\documentclass{iau}
\usepackage{graphicx}

\title[What Shapes Supernova Remnants?] 
{What Shapes Supernova Remnants?}

\author[Laura A. Lopez]   
{Laura A. Lopez$^1$
}

\affiliation{$^1$MIT-Kavli Institute for Astrophysics and Space Research, \\ 77 Massachusetts Ave., 37-664H, Cambridge, MA 02139 \\ email: {\tt lopez@space.mit.edu} \\[\affilskip]
}

\pubyear{2013}
\volume{296}  
\pagerange{}
\jname{Supernova Environmental Impacts}
\editors{A.~K. Ray \& R. McCray, eds.}
\begin{document}

\maketitle

\begin{abstract}
Evidence has mounted that Type Ia and core-collapse (CC) supernovae (SNe) can have substantial deviations from spherical symmetry; one such piece of evidence is the complex morphologies of supernova remnants (SNRs). However, the relative role of the explosion geometry and the environment in shaping SNRs remains an outstanding question. Recently, we have developed techniques to quantify the morphologies of SNRs, and we have applied these methods to the extensive X-ray and infrared archival images available of Milky Way and Magellanic Cloud SNRs. In this proceeding, we highlight some results from these studies, with particular emphasis on SNR asymmetries and whether they arise from ``nature'' or ``nurture''. 

\end{abstract}

\firstsection 
       \firstsection        
              
\section{Introduction}

In the past decades, evidence has mounted that supernova explosions (SNe) can have significant deviations from spherical symmetry. Spectropolarimetry studies -- the measure of the polarization of light as a function of wavelength as it is scattered through the debris layers of expanding SNe -- demonstrate that both Type Ia and core-collapse (CC) SNe are aspherical near maximum brightness (e.g., \cite[Wang \& Wheeler 2008]{wang08}; \cite[Kasen et al. 2009]{kasen09}). Line profiles in nebular spectra (100--200 days after explosion) of SNe show similar evidence of these ejecta asymmetries (e.g., \cite[Mazzali et al. 2001]{mazzali01}; \cite[Maeda et al. 2010]{maeda10}); confirmation of SN asymmetries is also possible at much later times via comparison of SN light echo spectra from different perspectives (e.g., \cite[Rest et al. 2011]{rest11}). Furthermore, proper motion studies demonstrate that pulsars have velocities up to $\sim$1000 km s$^{-1}$ (\cite[Lyne \& Lorimer 1994]{lyne94}; \cite[Faucher-Gigu{\`e}re \& Kaspi 2006]{faucher}), consistent with ``kicks'' imparted to newly-forming neutron stars in asymmetric SNe. The asymmetries may be inherent to the explosion mechanisms, and several mechanisms have been proposed that reproduce the necessary degrees of asymmetry: e.g., acoustic power generated in the proto-neutron star (\cite[Burrows et al. 2007]{burrows07}), the non-spherically symmetric standing accretion shock instability (\cite[Blondin \& Mezzacappa 2007]{blondin07}), and neutrino heating (\cite[Scheck et al. 2004]{scheck04}). 

Supernova remnants (SNRs) retain imprints from the geometry of their progenitors' explosions as well. For example, the three-dimensional projections of Cassiopeia~A reveal complex, asymmetric features which are attributed to the explosion \cite[(DeLaney et al. 2010)]{delaney10}. However, SNRs are also shaped by their environments; in particular, inhomogeneities in the circumstellar medium (CSM) structure affects the SNR morphology as well (e.g., Kepler's SNR: \cite[Reynolds et al. 2007]{reynolds07}). The relative role of the explosion and environment in shaping SNRs remains an outstanding question, and it has even been cited as the biggest challenge in modern SNR research (\cite[Canizares 2004]{canizares04}). 

Recently, we have developed techniques to quantify the morphological properties of SNRs (\cite[Lopez et al. 2009a]{lopez09a}), and we have applied these methods to archival {\it Chandra X-ray Observatory} and {\it Spitzer Space Telescope} images to assess the influence of ``nature'' versus ``nurture'' in SNR dynamics and evolution. In this proceeding, we highlight some results from these morphological studies to characterize SNR asymmetries. 

\section{Method}

The method we employed to characterize the symmetry of SNRs is a power-ratio method (PRM). The PRM has been used extensively to quantify the morphologies of galaxy clusters (\cite[Buote \& Tsai 1995, 1996; Jeltema et al. 2005]{b95,b96,jeltema05}). We extended the technique to compare the distribution of elements in individual SNRs (\cite[Lopez et al. 2009a]{lopez09a}) and to examine the symmetry of X-ray and IR emission in Type Ia and CC SNRs (\cite[Lopez et al. 2009b; Lopez et al. 2011; Peters et al. 2013]{lopez09b,lopez11,peters13}). We refer the reader to these papers for a detailed derivation and description of the method; we provide a basic summary below. 

The PRM measures asymmetries via calculation of the multipole moments of emission in a circular aperture. It is derived similarly to the expansion of a two-dimensional gravitational potential, except an image's surface brightness replaces the mass surface density. The powers $P_{\rm m}$ of the expansion are obtained by integrating the magnitude of each term $\Psi_{\rm m}$ over the aperture radius $R$. We divide the powers $P_{\rm m}$ by $P_{0}$ to normalize with respect to flux, and we set the origin position in our apertures to the centroids of our images so that the dipole power $P_{1}$ approaches zero. In this case, the higher-order terms reflect the asymmetries at successively smaller scales. The quadrupole power ratio $P_{2}/P_{0}$  reflects the ellipticity or elongation of a source, and the octupole power ratio $P_{3}/P_{0}$ quantifies the mirror asymmetry of a source.

\section{Results}

We applied the PRM to {\it Chandra} Advanced CCD Imaging Spectrometer (ACIS) and {\it Spitzer} Multiband Imaging Photometer (MIPS) images of Milky Way, LMC, and SMC SNRs, and we highlight the results and implications of these analyses below. 

\begin{figure}[ht]
\begin{center}
 \includegraphics[width=0.9\textwidth]{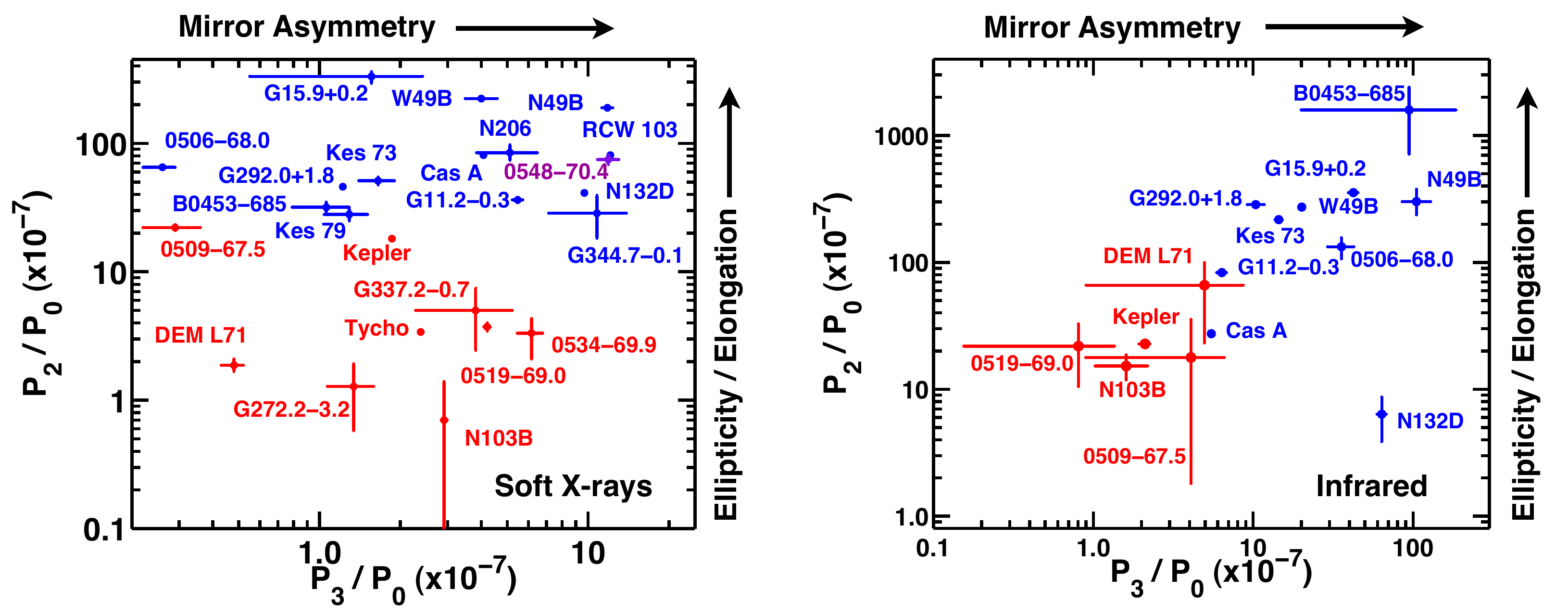} 
 \caption{Results from application of the power-ratio method to 24 galactic and LMC SNRs: quadrupole power ratio $P_{2}/P_{0}$ (which measures ellipticity/elongation) vs. octupole power ratio $P_{3}/P_{0}$ (which quantifies mirror asymmetry) of the soft X-ray band (0.5--2.1 keV) images (left) and of the {\it Spitzer} 24 $\mu$m images (right). Type Ia SNRs are plotted in red, and the CC SNRs are in blue (classified by abundance ratios). One source, 0548$-$70.4, is in purple because of its anomalous abundance ratios. The Type Ia SNRs separate naturally from the CC SNRs in this diagram. Figures are adapted from \cite[Lopez et al. (2011)]{lopez11} and \cite[Peters et al. (2013)]{peters13}.}
   \label{fig1}
\end{center}
\end{figure}

{\underline{\it Type Ia and CC SNRs have distinct symmetries}}. Using the PRM on X-ray line images of seventeen SNRs, we found that the emission from shock-heated metals in Type Ia SNRs is more spherical and more mirror symmetric than in CC SNRs (\cite[Lopez et al. 2009b]{lopez09b}). Furthermore, this result holds for the X-ray bremsstrahlung emission of Type Ia and CC SNRs as well (see Figure~\ref{fig1}; \cite[Lopez et al. 2011]{lopez11}). The ability to distinguish between the two SN classes based on bremsstrahlung emission morphology alone suggests the potential to type SNRs with weak X-ray lines or with low spectral resolution X-ray observations. 

\begin{figure}[ht]
\begin{center}
 \includegraphics[width=\textwidth]{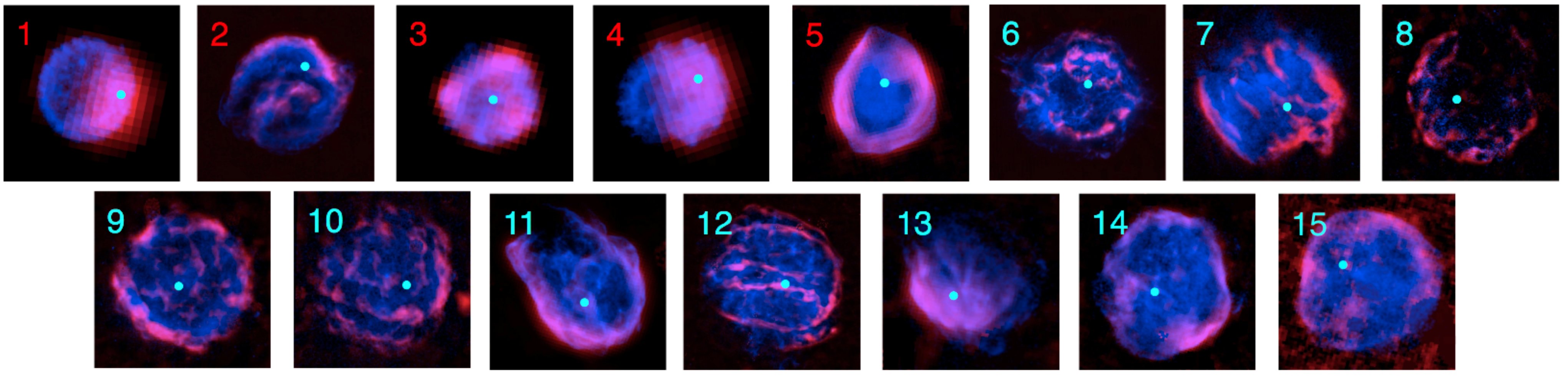} 
 \caption{Images (adapted from \cite[Peters et al. 2013]{peters13}) of the {\it Spitzer} 24 $\mu$m emission (red) and {\it Chandra} soft X-ray (0.5--2.1 keV) emission of the 15 SNRs analyzed in Figure~\ref{fig1} (right). The thermal X-rays trace the ejecta material heated by the reverse shock, while the IR originates from circumstellar dust heated by interaction with the blast wave. Red numbers denote Type Ia SNRs; blue numbers denote CC SNRs. SNRs are as follows: [1] 0509$-$67.5; [2] Kepler; [3] 0519--69.0; [4] N103B; [5] DEM~L71; [6] Cas~A; [7] W49B; [8] G15.9$+$0.2; [9] G11.2$-$0.3; [10] Kes~73; [11] N132D; [12] G292.0$+$1.8; [13] 0506$-$68.0; [14] N49B; [15] B0453$-$685.}
   \label{fig1_charee}
\end{center}
\end{figure}

We further extended the technique to 24 $\mu$m {\it Spitzer} images of the same sample, and we found that the warm dust emission of Type Ia and CC SNRs also has distinct symmetries (\cite[Peters et al. 2013; see Fig.~\ref{fig1}]{peters13}). Similar to the X-ray results, Type Ia SNRs have more circular and mirror symmetric IR emission than CC SNRs. The two wavelength regimes probe distinct emitting regions (see Figure~\ref{fig1_charee}): the thermal X-rays trace the ejecta material heated by the reverse shock, while the IR originates from circumstellar dust heated by interaction with the blast wave. Yet both are sensitive to the shape of the contact discontinuity, where the ejecta are impacting the shocked ISM. As the contact discontinuity is shaped by the explosion geometry as well as the structure of the surrounding medium, the distinct symmetries of Type Ia and CC SNRs in the X-ray and IR wavebands reflect both the different explosion mechanisms and the environments of the two classes. 

It is noteworthy that SNRs with bright neutron stars/pulsars (e.g., G11.2$-$0.3, Kes 73, B0453$-$685) are among the most circular of the CC SNRs. Based on our preliminary investigations, CC SNRs with neutron stars appear more spherical and symmetric than CC SNRs without neutron stars. However, strict limits on the presence/absence of neutron stars in SNRs are necessary before the trend can be established statistically. 

\begin{figure}[t]
\begin{center}
 \includegraphics[width=0.85\textwidth]{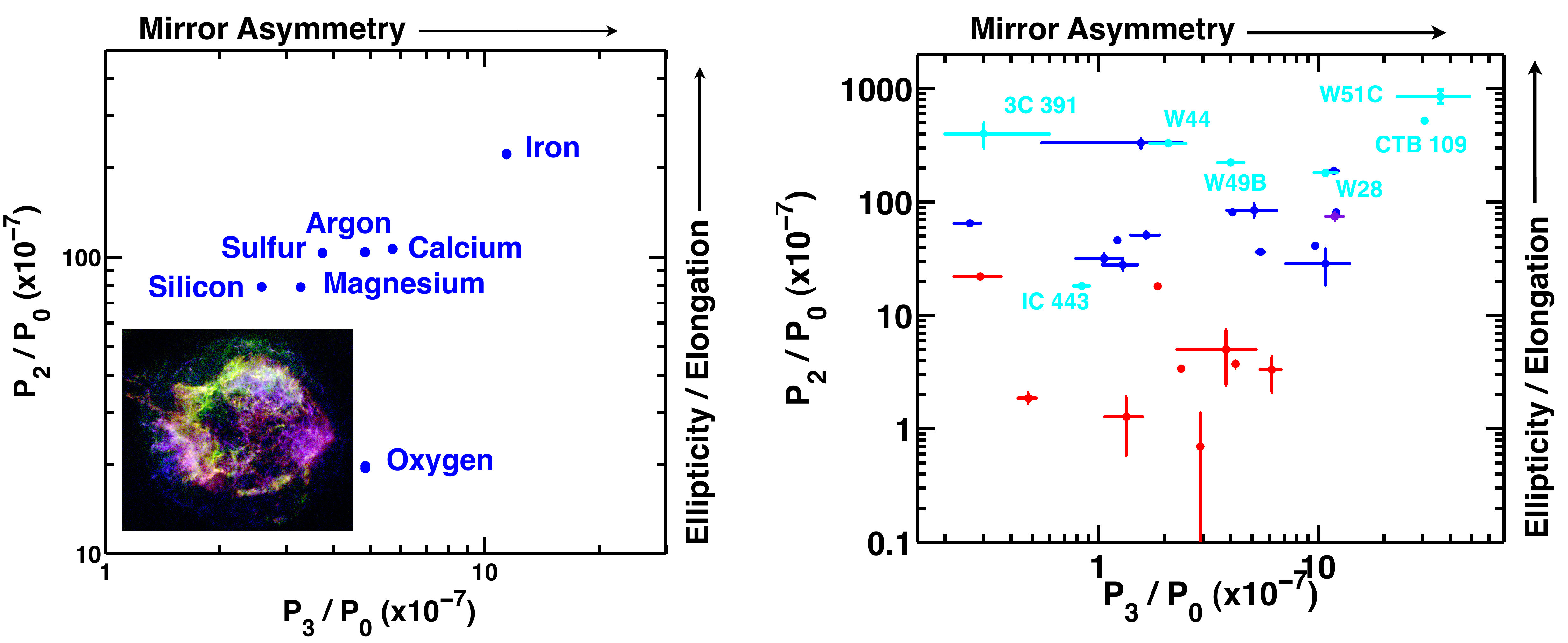} 
 \caption{{\it Left}: Quadrupole power ratio $P_{2}/P_{0}$ vs. octupole power ratio $P_{3}/P_{0}$ for 7 X-ray line images of Cassiopeia~A (O {\sc viii}, Mg {\sc xi}, Si {\sc xiii}, S {\sc xv}, Ar {\sc xvii}, Ca {\sc xix}, and Fe {\sc xxv}). The inset image shows the distribution of O {\sc viii} (green), Si {\sc xiii} (red), and Fe {\sc xxv} (blue). Plot and images were produced using data in \cite[Lopez et al. 2011]{lopez11}). {\it Right}: $P_{2}/P_{0}$ vs. $P_{3}/P_{0}$ for SNRs interacting with molecular clouds (in cyan); non-interacting SNRs are from Fig.~\ref{fig1}, with Type Ia SNRs in red and CC SNRs in blue. Interacting SNRs are more elliptical than non-interacting SNRs.}
   \label{fig2}
\end{center}
\end{figure}

{\underline{\it Elements within individual SNRs have distinct symmetries}}. Our quantitative approach also enables us to compare images of different emission features (like the morphologies of the shock-heated metals) within individual remnants. For example, Figure~\ref{fig2} (left) plots the symmetry diagram for seven X-ray emission line images of Cassiopeia~A (O {\sc viii}, Mg {\sc xi}, Si {\sc xiii}, S {\sc xv}, Ar {\sc xvii}, Ca {\sc xix}, and Fe {\sc xxv}). The O has the most symmetric distribution, while the Fe is comparatively more asymmetric and elongated than the other metals. By contrast, the intermediate-mass elements (Mg, Si, S, Ar, and Ca) appear to have similar morphologies. Thus, despite relatively efficient mixing of the ejecta (\cite[Lopez et al. 2011]{lopez11}), the distinct symmetries of the metals reflect imprints of the explosion geometry. 

{\underline{\it Environment Shapes Large-Scale SNR Morphology}}. To examine the role of environment in shaping SNRs,  we have compared systematically the morphological properties of SNRs in different ISM conditions. For example, we have measured the symmetry of SNRs thought to be interacting with molecular clouds (see Figure~\ref{fig2}, right), based on coincidence with OH masers (which indicate the presence of shocked, molecular gas: \cite[Wardle \& Yusef-Zadeh 2002]{wardle02}). The interacting SNRs are the most elliptical of the CC SNRs, evidence that environment has a dramatic large-scale effect on SNR morphologies.

\section{Conclusions}

We have begun to address the ``nature'' vs. ``nurture'' conundrum in SNR science using a systematic approach on available archival data. Many exciting questions and prospects remain, including application of the techniques to other wavelength images, investigation of SNRs in other galaxies, characterization of SNR morphological evolution with age, and comparison to hydrodynamical model predictions.


\end{document}